\documentclass[conference]{IEEEtran}

\usepackage{cite}
\usepackage{amsmath,amssymb,amsfonts}
\usepackage{graphicx}
\usepackage{textcomp}
\usepackage{xcolor}
\usepackage{listings}
\usepackage{url}
\usepackage{hyperref}
\usepackage{booktabs}
\usepackage{tikz}
\usetikzlibrary{shapes.geometric, arrows.meta, positioning, fit, backgrounds, calc}

\def\BibTeX{{\rm B\kern-.05em{\sc i\kern-.025em b}\kern-.08em
    T\kern-.1667em\lower.7ex\hbox{E}\kern-.125emX}}

\begin{document}

\title{AI-Native Network Controller: A Modular Framework for Safe Agentic Control of Multi-Domain Network Infrastructure}

\author{\IEEEauthorblockN{Merim Dzaferagic}
\IEEEauthorblockA{School of Computer Science and Statistics, Trinity College Dublin, Ireland \\
merim.dzaferagic@tcd.ie}}

\maketitle

\begin{abstract}
The convergence of multiple network domains, including radio access, optical transport, and core networks, under unified intelligent control is a fundamental requirement for future 6G systems. This is important because existing network controllers remain largely domain-specific, such as the O-RAN RIC for radio, or they lack native support for AI-driven automation across heterogeneous infrastructure. As a result, safe and coordinated agentic control of multi-domain networks is still an open challenge. In this paper, we present the AI-Native Network Controller (AI-NNC), an open-source and modular framework that enables agentic AI control across diverse network domains. The framework is designed around a protocol-agnostic architecture in which each physical device is integrated through a lightweight Python adapter, while control logic is implemented through domain-specific control applications. 
% The main contributions of this work are threefold: (i) a unified node integration layer that bridges any network element to the controller through a ZeroMQ-based message broker and registration plane, (ii) an application framework in which control applications act as validated tools that AI agents can safely invoke, and (iii) a command validation mechanism that enforces safety constraints and provides guarantees for AI-controlled critical infrastructure. 
Beyond closed-loop control, the framework also supports dataset collection, agentic AI experimentation, and coordinated testbed operation using the same validated control and measurement interfaces. This design enables a safer paradigm for autonomous network management, where AI agents operate through validated applications rather than issuing commands directly to network equipment.
\end{abstract}


%======================================================================
\section{Introduction}
\label{sec:introduction}
%======================================================================
Next-generation 6G systems require unified intelligent control across multiple network domains, including radio access networks (RAN), optical transport, and core infrastructure. In this paper, we present the \emph{AI-Native Network Controller (AI-NNC)}, an open-source modular framework designed to enable safe agentic AI control over heterogeneous multi-domain network environments.

Such a unified control approach is important because current network infrastructures remain fragmented across domains, protocols, and vendor-specific management systems, making coordinated automation difficult to achieve at scale~\cite{foukas2017slicing, afolabi2018slicing}. The transition toward 6G further intensifies these challenges by introducing stringent requirements for cross-domain orchestration, autonomous operation, and ultra-low latency services~\cite{ebrahimi2024resource}. Without an AI-native control plane that spans all network segments, the vision of fully autonomous multi-domain network management remains unrealized.

The AI-NNC addresses this gap through a protocol-agnostic architecture that decouples control logic from device-specific interfaces. Network elements are integrated through lightweight Python adapters, while domain-specific control applications provide structured tools that AI agents can invoke safely. Inspired by the architectural principles identified in our prior work on decentralized AI control for multi-party multi-network 6G deployments~\cite{dzaferagic2025icc}, the AI-NNC offers a concrete and deployable realization of these concepts.

Beyond closed-loop network control, the design of the AI-NNC enables a broader range of experimental and operational use cases. Since all network nodes expose measurements and control functions through a uniform interface, the framework can be used to collect datasets from multi-domain testbed experiments without executing any control actions, by simply subscribing to measurements and exporting them to external storage. Similarly, control applications can be used as lightweight experimentation or orchestration scripts that configure network nodes, coordinate experiment execution, and monitor system behavior. These capabilities allow the AI-NNC to function not only as a network controller, but also as a unified experimentation and coordination framework for multi-domain network research.

A central motivation for this design is the emergence of agentic AI systems that can perceive network state, reason autonomously, and execute control actions~\cite{abou2025agentic, li2025agentic}. While such agents offer transformative opportunities for network optimization, directly exposing critical infrastructure to autonomous decision-making introduces unacceptable safety risks. The AI-NNC therefore enforces a layered safety model where agents interact only through validated control applications, supported by explicit command validation mechanisms grounded in safe reinforcement learning principles~\cite{garcia2015safe, alshiekh2018shielding}.

The main contributions of this paper are:
\begin{itemize}
    \item A modular and protocol-agnostic controller architecture in which integrating a new network node requires only a single lightweight Python adapter script, enabling rapid deployment in both operational and experimental testbed environments.
    \item A control application framework where domain-specific applications, whether classical or AI-based, serve as validated tools for agentic AI through declarative Python definitions.
    \item A command validation mechanism that enforces safety guarantees for AI-controlled critical infrastructure, enabling conflict detection between applications and safe autonomous operation.
\end{itemize}

%======================================================================
\section{Related Work}
\label{sec:related}
%======================================================================

\subsection{SDN and Multi-Domain Network Controllers}

Software-defined networking (SDN) introduced the foundational principle of separating control logic from forwarding behavior through centralized programmable controllers~\cite{mckeown2008openflow, kreutz2015sdn}. This paradigm enabled flexible control of packet-switched infrastructures and motivated the development of widely adopted platforms such as OpenDaylight~\cite{medved2014opendaylight} and ONOS~\cite{berde2014onos}. These controllers provide extensible southbound protocol support and model-driven architectures, but they were primarily designed for transport and packet domains rather than end-to-end multi-domain orchestration.

In parallel, the O-RAN Alliance extended the SDN concept into the radio domain through the RAN Intelligent Controller (RIC), which hosts modular xApps for near-real-time control and rApps for longer-term optimization~\cite{polese2023oran, bonati2021intelligence}. While O-RAN represents a major step toward intelligent and modular RAN automation, its scope remains focused on radio access networks and does not directly address unified control of optical transport or core network functions. Recent work highlights cross-domain interoperability as one of the most important unresolved challenges for O-RAN-based deployments~\cite{santos2025managing}.

Overall, existing controller ecosystems remain domain-specific, leaving an open gap for architectures that can provide unified AI-native control across heterogeneous network segments.

\subsection{AI and Machine Learning for Network Automation}

AI-driven automation has become a major research direction in telecommunications, ranging from early deep reinforcement learning approaches for resource allocation and network optimization~\cite{mao2016resource} to more recent developments in large language model (LLM)-based network operations~\cite{zhou2024llm, hong2025llm}. These advances suggest significant potential for autonomous control systems that operate beyond the capabilities of manual network management.

Standardization bodies have also introduced architectural frameworks for closed-loop automation. In particular, 3GPP's Network Data Analytics Function (NWDAF) defines analytics-driven control support within the 5G core~\cite{3gpp23288}, while ETSI's Zero-touch Service Management (ZSM) architecture outlines principles for fully automated service and network management~\cite{etsizsm002}. However, such frameworks typically prescribe structured integration patterns and do not provide a flexible, protocol-agnostic control layer that spans diverse physical domains.

\subsection{Agentic AI and Safety in Network Management}

The concept of agentic AI, where autonomous agents plan, coordinate, and execute operational decisions, is gaining increasing attention in network research~\cite{li2025agentic, demirel2026intents}. The Agentic-AI Core (A-Core) framework proposes a mission-oriented approach in which specialized agents manage network services as goal-driven tasks~\cite{li2025agentic}. Similarly, surveys of agentic architectures emphasize their potential for autonomous infrastructure operation~\cite{abou2025agentic}.

Despite these developments, practical safety mechanisms remain a critical missing component. Theoretical foundations such as safe reinforcement learning~\cite{garcia2015safe} and shielding-based enforcement~\cite{alshiekh2018shielding} provide important models for constraining unsafe actions, but they have not yet been integrated into deployable multi-domain network controllers.

Our prior work established the architectural foundations for decentralized and multi-party AI control across heterogeneous 6G environments~\cite{dzaferagic2025icc, dzaferagic2025modular}. The AI-NNC presented in this paper builds directly on these principles by providing an open-source, modular implementation that combines agentic control with explicit validation and safety enforcement mechanisms.

%======================================================================
\section{Architecture}
\label{sec:architecture}
%======================================================================

\begin{figure}[t]
\centering
\begin{tikzpicture}[
    node distance=0.4cm and 0.3cm,
    box/.style={rectangle, draw, rounded corners=2pt, minimum width=2.1cm, minimum height=0.55cm, font=\scriptsize, align=center, fill=#1},
    box/.default=blue!8,
    smallbox/.style={rectangle, draw, rounded corners=2pt, minimum width=1.4cm, minimum height=0.45cm, font=\tiny, align=center, fill=#1},
    smallbox/.default=green!10,
    nodebox/.style={rectangle, draw, rounded corners=2pt, minimum width=1.1cm, minimum height=0.45cm, font=\tiny, align=center, fill=orange!15},
    plane/.style={rectangle, draw, dashed, rounded corners=4pt, inner sep=4pt, fill=#1},
    plane/.default=blue!3,
    arrow/.style={-{Stealth[length=2pt]}, thick},
    biarrow/.style={{Stealth[length=2pt]}-{Stealth[length=2pt]}, thick},
    label/.style={font=\tiny\bfseries, fill=white, inner sep=1pt}
]

% Control Plane
\node[box=blue!12] (server) {FastAPI\\Server};
\node[box=blue!12, right=0.3cm of server] (mcp) {MCP\\Server};
\node[box=purple!12, below=0.3cm of server] (app1) {App 1};
\node[box=purple!12, right=0.2cm of app1] (app2) {App 2};
\node[box=purple!12, right=0.2cm of app2] (app3) {App N};
\node[box=red!12, below=0.3cm of app2] (validator) {Command\\Validator};

\begin{scope}[on background layer]
\node[plane=blue!5, fit=(server)(mcp)(app1)(app3)(validator), label=above:{\tiny\bfseries Control Plane}] (cp) {};
\end{scope}

% AI Agent
\node[box=yellow!25, above=0.5cm of mcp] (agent) {AI Agent\\(LLM)};

% Registration Plane
\node[box=green!15, below=0.9cm of validator] (register) {Register\\Service};
\node[box=green!15, right=0.3cm of register] (redis) {Redis\\Store};

\begin{scope}[on background layer]
\node[plane=green!5, fit=(register)(redis), label=above:{\tiny\bfseries Registration Plane}] (rp) {};
\end{scope}

% Data Plane
\node[box=cyan!15, below=0.9cm of register] (broker) {ZMQ Message\\Broker};

\begin{scope}[on background layer]
\node[plane=cyan!5, fit=(broker), inner sep=6pt, minimum width=4.5cm, label=above:{\tiny\bfseries Data Plane}] (dp) {};
\end{scope}

% Network Nodes
\node[nodebox, below=0.8cm of broker, xshift=-1.2cm] (n1) {Optical\\Node};
\node[nodebox, right=0.15cm of n1] (n2) {RAN\\Node};
\node[nodebox, right=0.15cm of n2] (n3) {Core\\Node};

\begin{scope}[on background layer]
\node[plane=orange!5, fit=(n1)(n3), inner sep=4pt, label=above:{\tiny\bfseries Network Nodes}] (nn) {};
\end{scope}

% Arrows
\draw[biarrow, blue!60] (agent) -- (mcp);
\draw[biarrow, blue!60] (server) -- (mcp);
\draw[arrow, purple!60] (app1.south) -- ++(0,-0.15) -| (validator.north west);
\draw[arrow, purple!60] (app2) -- (validator);
\draw[arrow, purple!60] (app3.south) -- ++(0,-0.15) -| (validator.north east);
\draw[biarrow, green!50!black] (validator.south) -- ++(0, -0.25) -| (register.north);
\draw[biarrow, green!50!black] (register) -- (redis);
\draw[biarrow, cyan!60!black] (register.south) -- ++(0,-0.25) -| (broker.north);
\draw[biarrow, orange!60!black] (broker) -- (n1.north -| broker.south west);
\draw[biarrow, orange!60!black] (broker) -- (n2);
\draw[biarrow, orange!60!black] (broker) -- (n3.north -| broker.south east);

\end{tikzpicture}
\caption{AI-NNC three-plane architecture. AI agents interact through the MCP server; control applications process measurements and issue validated commands; network nodes communicate through the ZMQ broker.}
\label{fig:architecture}
\end{figure}
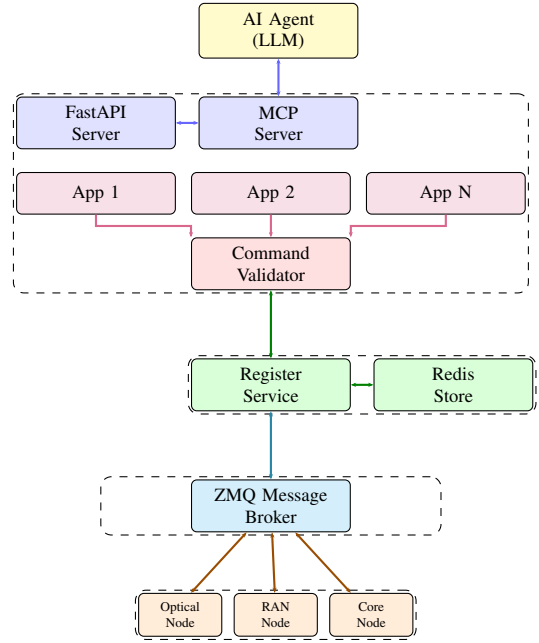

The AI-NNC is structured as a three-plane architecture consisting of a control plane, a registration plane, and a data plane, connected through ZeroMQ~\cite{zmq} message passing and backed by Redis for persistent state (Fig.~\ref{fig:architecture}). In addition, the architecture is designed to seamlessly integrate with external data storage and processing systems, allowing measurements and control-related data to be streamed or exported to databases and data pipelines such as time-series databases (e.g., ModelarDB), object storage systems (e.g., MinIO or other S3-compatible storage), and downstream analytics or machine learning pipelines without requiring changes to the core framework.

\subsection{Data Plane: Protocol-Agnostic Message Broker}

The data plane employs a ZeroMQ-based message broker that provides bidirectional communication between network nodes and control applications through four ports using PUSH/PULL and PUB/SUB patterns. Nodes push measurements to the broker, which publishes them to subscribed applications via topic-based filtering. Commands flow in reverse: applications push commands to the broker, which publishes them to the target nodes. This design fully decouples applications from nodes, since neither side needs to know the other's protocol, address, or implementation.

The message format is uniform across all domains: a topic identifier (node ID) followed by a JSON payload. This simplicity is deliberate, as standardizing only the message envelope allows the framework to accommodate any measurement type or command structure without requiring modifications to the broker.

\subsection{Registration Plane: Discovery and State}

The registration plane provides a REQ/REP-based service where nodes and applications register their capabilities. Each network node registers its available \textit{performance metrics} (PMs) and \textit{control functions}. Each control application registers which PMs it reads and which control functions it invokes. Redis provides persistent storage for this metadata, enabling dynamic discovery and runtime validation.

This plane serves a dual purpose. First, it supports operational functions such as routing and subscription management. Second, it provides governance by tracking which applications control which nodes, thereby enabling conflict detection.

\subsection{Control Plane: Applications and AI Integration}

The control plane hosts the FastAPI server, the MCP (Model Context Protocol) server for AI agent integration, and all control applications. Each application runs in its own thread with an independent control loop, receiving filtered measurements and issuing commands through the validation pipeline.

\subsection{Network Node Integration}
\label{sec:node_integration}

A central design goal is minimal integration effort for new network equipment. Integrating any device, regardless of its native protocol (REST API, NETCONF, SNMP, gRPC, or proprietary), requires only a single Python adapter script that implements a simple contract:

\begin{enumerate}
    \item \textbf{Register}: Automatically handled based on the \texttt{node.conf} file. It takes all \texttt{available\_measurements} and \texttt{available\_controls} and registers them with the register. 
    \item \textbf{Setup}: The initial setup of the node. This is optional and allows for initial credential setup for external access to databases (e.g. influxDB, Prometheus). 
    \item \textbf{Poll Measurements}: The southbound interface is polled to retrieve the most recent measurement values from the network nodes that is being controlled. This operation is executed periodically every \texttt{measurement\_interval} seconds by the \texttt{NodeRunner} component. At each cycle, the function returns a dictionary mapping measurement names to their corresponding values when data is available; otherwise, it returns \texttt{None} to indicate that no measurements were obtained during that interval.
    \item \textbf{Handle Commands}: Incoming control commands from an \texttt{aic\_app} are received through the message bus and processed by this method. Nodes that support actuation or configuration updates should override this handler, whereas measurement-only nodes are not required to implement it. The command payload is provided as a dictionary specifying the target control parameters, and the method returns a boolean value indicating whether the command was successfully applied.
\end{enumerate}

Listing~\ref{lst:node} shows a minimal node adapter. The adapter translates between the controller's uniform ZMQ interface and the device's native protocol. This bridge pattern ensures that the framework does not require modification when supporting new device types, since all protocol-specific logic is encapsulated within the adapter. For example, the framework's existing optical node adapters, wrap REST API calls behind this pattern, while dummy nodes generate simulated measurements and present identical interfaces to the controller. The same approach has been applied in the radio domain through an adapter implemented for an \texttt{srsRAN} gNB, targeting the O-DU, where O-DU telemetry and control parameters are bridged to the controller using the same uniform ZMQ interface. Adding a core network function, or any other device follows the same pattern.

\begin{lstlisting}[caption={Minimal network node adapter (simplified). The adapter bridges the controller's ZMQ interface to the device's native REST API.\\}, label=lst:node, float=t]
  @node(name="DummyNode")
  class NetworkNode(ControlledEntity):
      available_measurements = ["dummy_value"]
      available_controls = ["SET_GAIN", "SET_VOA", "SET_TILT"]
      measurement_interval = 1.0

      def setup(self):
          self._dummy_value = 0.0

      def poll_measurements(self):
          return {"dummy_value": self._dummy_value}

      def handle_command(self, payload):
          # Expect payload like {"dummy_value": 42.0}
          if "dummy_value" not in payload:
              return False
          try:
              self._dummy_value = float(payload["dummy_value"])
          except (TypeError, ValueError):
              return False
          return True

  if __name__ == "__main__":
      NodeRunner().run()
\end{lstlisting}

%======================================================================
\section{Control Applications as Agent Tools}
\label{sec:applications}
%======================================================================

Control applications are the central abstraction in the AI-NNC. They are not merely scripts that react to network events; they are \textit{validated, discoverable tools} that AI agents can invoke to affect the network. This section describes how applications are defined, how they enable agentic control, and how the validation layer ensures safety.

\subsection{Declarative Application Definition}

Applications are defined using Python decorators that declare their measurement subscriptions, control functions, and processing behavior:

\begin{lstlisting}[caption={Control application definition using decorators. Each application declares what it reads and what it controls.\\}, label=lst:app, float=t]
@aic_app(name="GainOptimizer")
class GainOptimizer(AicApp):
    aic_app_id = 1
    control_loop_update_time = 2  # seconds

    read_measurements = {
        3: ["power_in", "power_out"],
        8: ["preamp_gain", "snr"]}

    control_functions = {
        8: ["SET_GAIN", "SET_VOA"]}

    @classmethod
    @command_validator("SET_GAIN")
    def validate_gain(cls, params):
        if params["target_gain"] > 25.0:
            return False, "Gain exceeds safe limit"
        if params["target_gain"] < 0:
            return False, "Gain below minimum"
        return True, None

    @classmethod
    @agent_controlled(
        name="optimize",
        description="Optimize gain for target",
        schema={"properties": {
            "node_id": {"type": "integer"},
            "strategy": {"type": "string"}}})
    def handle_optimize(cls, req, measurements):
        current = measurements[8][-1]["snr"]
        # Compute optimal gain ...
        cls.add_command(("SET_GAIN", {"node_id": 8, "value": {"amp_type": "preamp", "target_gain": new_gain}}))
        return {"applied_gain": new_gain}

    def process(self, measurements):
        # Classical or AI control logic
        for m in measurements.get(8, []):
            if m["snr"] < threshold:
                cls.add_command(("SET_GAIN", {"node_id": 8, "value": {"amp_type": "preamp", "target_gain": new_gain}}))
\end{lstlisting}

\textbf{The \texttt{@aic\_app} decorator automatically generates REST API endpoints and MCP tools for the application, including tools for querying measurements, changing application state, invoking control functions, and calling agent-controlled methods.} As a result, adding a new control application to the system, even one that operates on a completely different network domain, requires no modification to the framework.

\subsection{Classical and AI Control}

Critically, control applications \textit{do not require AI}. The \texttt{process()} method can implement any control logic, including PID controllers, threshold-based rules, optimization algorithms, or deep reinforcement learning agents. This flexibility is essential for practical deployment, since proven classical algorithms may be preferred for certain control loops, while AI-based approaches are applied to others.

When AI is used, the \texttt{@agent\_controlled} decorator enables LLM-based agents to invoke application logic within the control loop while accessing live measurements. This differs fundamentally from exposing raw network APIs directly to an AI agent, as the agent instead operates through a domain-specific application that encapsulates expert knowledge, safety constraints, and validated command execution.

\subsection{MCP Integration for Agentic Control}

The Model Context Protocol (MCP) server automatically generates tools from all registered applications. An AI agent (e.g., \texttt{Claude, ChatGPT, OpenCode, OpenClaw}) can discover available tools, query measurements, invoke control functions, and execute agent-controlled operations through a standardized JSON-RPC interface. The tool generation is fully automatic:

\begin{itemize}
    \item Each \textbf{control function} becomes an MCP tool with validated input schemas.
    \item Each \textbf{measurement subscription} becomes a query tool.
    \item Each \textbf{@agent\_controlled method} becomes an invocable tool that executes within the application's process loop with access to live state.
    \item \textbf{Application state management} (start, pause, stop) is exposed as tools.
\end{itemize}

This design ensures that AI agents interact with the network through the same validated paths as automated control loops, with no special privileges or bypass mechanisms. \textbf{This interaction model naturally supports intent-based networking, where high-level objectives expressed by an AI agent are translated into validated control actions through domain-specific applications rather than direct device commands.}

%======================================================================
\section{Command Validation for Safe Agentic Control}
\label{sec:validation}
%======================================================================

The command validation mechanism is the critical safety layer that distinguishes the AI-NNC from approaches that expose network APIs directly to AI agents. Its importance is amplified in the context of critical infrastructure, including radio networks, optical transport, and core network functions, where incorrect commands can cause service outages affecting millions of users.

\subsection{Validation Architecture}

\begin{figure}[t]
\centering
\begin{tikzpicture}[
    node distance=0.35cm,
    box/.style={rectangle, draw, rounded corners=2pt, minimum width=2.5cm, minimum height=0.5cm, font=\scriptsize, align=center, fill=#1},
    box/.default=white,
    decision/.style={diamond, draw, aspect=2, inner sep=1pt, font=\scriptsize, fill=yellow!15},
    arrow/.style={-{Stealth[length=2pt]}, thick},
    label/.style={font=\tiny, fill=white, inner sep=1pt}
]

\node[box=blue!10] (agent) {AI Agent (MCP)};
\node[box=purple!10, right=0.4cm of agent] (process) {App process()};
\node[box=green!10, left=0.4cm of agent] (rest) {REST API};

\node[box=orange!10, below=0.5cm of agent] (queue) {Command Queue};

\node[decision, below=0.5cm of queue] (validate) {Valid?};

\node[box=red!10, left=0.8cm of validate] (reject) {Reject +\\Log};
\node[box=green!15, below=0.5cm of validate] (execute) {Execute\\Command};
\node[box=cyan!10, below=0.35cm of execute] (broker) {ZMQ Broker};
\node[box=orange!10, below=0.35cm of broker] (node) {Network Node};

\draw[arrow] (rest) |- (queue);
\draw[arrow] (agent) -- (queue);
\draw[arrow] (process) |- (queue);
\draw[arrow] (queue) -- (validate);
\draw[arrow] (validate) -- node[label, left] {\tiny No} (reject);
\draw[arrow] (validate) -- node[label, right] {\tiny Yes} (execute);
\draw[arrow] (execute) -- (broker);
\draw[arrow] (broker) -- (node);

\end{tikzpicture}
\caption{Unified command validation pipeline. All command sources, including AI agents, automated process loops, and manual REST API calls, pass through identical validation before reaching network equipment.}
\vspace{-1em}
\label{fig:validation}
\end{figure}
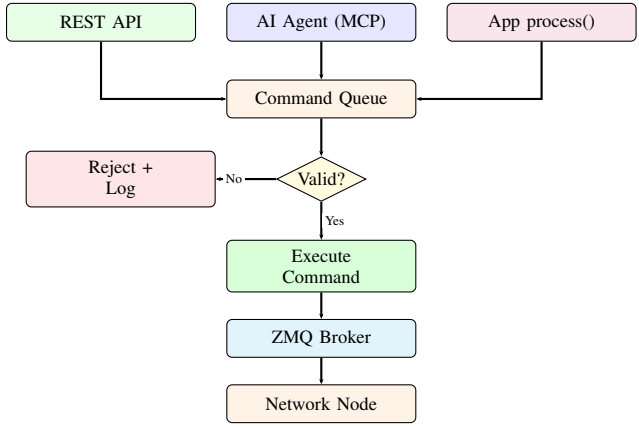

The \texttt{@command\_validator} decorator associates validation functions with specific command types. Every command, whether originating from an application's \texttt{process()} loop, a REST API call, or an AI agent via MCP, passes through the same validation pipeline (Fig.~\ref{fig:validation}).

Validators enforce domain-specific safety constraints. For optical networks, these include gain limits that prevent amplifier damage, wavelength boundaries, and power thresholds. For radio networks, validators could enforce transmission power limits, frequency allocation rules, and interference constraints. Each validator receives the command parameters and returns a boolean verdict with an optional rejection reason, providing explainability for both human operators and AI agents.

\subsection{Safety Guarantees for Critical Infrastructure}

The validation layer provides several properties essential for deploying AI control in critical networks:

\textbf{Uniform enforcement.} The same validation applies regardless of command origin. An AI agent cannot bypass safety checks that apply to automated control loops.

\textbf{Domain-specific constraints.} Validators encode equipment-specific safety limits derived from physical properties (e.g., maximum amplifier gain before saturation, minimum channel spacing to avoid crosstalk). These constraints are defined by domain experts alongside the control applications.

\textbf{Auditability.} All command attempts (valid and rejected) are logged with their source, parameters, and validation result. This provides the audit trail required for critical infrastructure compliance.

\textbf{Compositional safety.} Because validators are attached to command types rather than command sources, adding new applications or AI agents does not require updating safety rules. The safety guarantees compose automatically.

\subsection{Conflict Detection Between Applications}

The registration plane tracks which applications control which nodes and which control functions they invoke. This metadata enables a critical capability: \textit{direct conflict detection between control applications}.

When multiple applications issue competing commands to the same node (e.g., one increasing gain while another decreases it), the framework can detect and resolve these conflicts. The AI-NNC supports dedicated conflict mitigation applications (demonstrated in the reference implementation as \texttt{ConflictMitigator}) that subscribe to all measurements, observe the behavior of other applications, and arbitrate competing commands. These mechanisms can be used to detect indirect and implicit conflicts. 

In an agentic AI context, this conflict detection becomes even more powerful. An AI agent can:
\begin{itemize}
    \item Query the registration plane to understand which applications are active and what they control.
    \item Detect conflicting objectives between applications through measurement analysis.
    \item Coordinate between applications by adjusting their states (pause, resume) or parameters.
    \item Reason about the global network state across domains, something individual applications cannot do.
\end{itemize}

This positions the AI agent not as a direct controller of network equipment, but as a \textit{meta-controller} that orchestrates domain-specific applications, which is a fundamentally safer paradigm for autonomous network management.

%======================================================================
\section{Discussion}
\label{sec:discussion}
%======================================================================

\subsection{Integration Effort}

Table~\ref{tab:comparison} compares the integration effort for adding new network domains to the AI-NNC versus existing approaches. The AI-NNC requires only a single Python adapter per node type and a single application definition per control domain, with no modifications to the core framework.

\begin{table}[t]
\centering
\caption{Integration effort comparison for adding a new network domain}
\label{tab:comparison}
\footnotesize
\begin{tabular}{@{}lcc@{}}
\toprule
\textbf{Component} & \textbf{AI-NNC} & \textbf{Traditional} \\
\midrule
Node adapter & 1 Python script & Protocol plugin \\
Control logic & 1 app class & Platform-specific \\
AI integration & Automatic (MCP) & Custom API \\
Safety rules & Decorator & Separate system \\
Framework changes & None & Often required \\
\bottomrule
\end{tabular}
\end{table}

\subsection{Comparison with O-RAN RIC}

The AI-NNC shares the O-RAN RIC's philosophy of modular, pluggable applications (xApps/rApps) but differs in scope and design. The O-RAN RIC is domain-specific to radio, uses E2/A1/O1 interfaces, and does not natively support LLM-based agentic control. The AI-NNC is domain-agnostic, uses a uniform ZMQ-based interface, and provides first-class MCP integration for AI agents. However, the two are complementary: an AI-NNC node adapter could bridge to an O-RAN Near-RT RIC via its A1 or E2 interfaces, enabling unified multi-domain control that includes the RAN domain~\cite{dzaferagic2025modular}.

\subsection{Enabling Safe Agentic AI}

The layered architecture, in which AI agents operate through validated applications rather than directly on network equipment, addresses key concerns related to agentic AI in critical infrastructure~\cite{dhs2024ai, nist2023ai}. Each layer contributes additional safety:

\begin{enumerate}
    \item \textbf{Application layer}: Domain expertise encoded in control logic and validated commands. The applications themselves are the tools that the AI agent uses.
    \item \textbf{Validation layer}: Physical safety constraints enforced uniformly, acting as a shield~\cite{alshiekh2018shielding} against unsafe actions regardless of their source.
    \item \textbf{Registration layer}: Visibility into the full control topology, enabling conflict detection and resolution.
\end{enumerate}

This design means that the ``tools'' used by AI agents are not raw network APIs but curated and validated control interfaces that provide confidence through domain-specific logic. The AI agent reasons at a higher level of abstraction by choosing strategies, coordinating applications, and resolving conflicts, while the safety-critical details of command execution remain within validated application code.

%======================================================================
\section{Use Cases Enabled by AI-NNC}
\label{sec:usecases}
%======================================================================

The modular and protocol-agnostic design of the AI-NNC enables a broad range of use cases beyond traditional closed-loop network control. By decoupling node integration, measurement collection, control logic, and AI interaction, the framework can be applied flexibly across operational, experimental, and research-driven scenarios.

\subsection{Multi-Domain Network Control}

The primary use case of the AI-NNC is as a network controller for heterogeneous multi-domain infrastructure. Network nodes register their performance metrics and control functions, while control applications implement domain-specific logic to process measurements and issue validated commands. This model applies uniformly across radio, optical, and core network elements, allowing coordinated control without requiring protocol-specific logic in the controller itself.

Control applications are implemented as lightweight Python scripts, which simplifies rapid development, testing, and deployment of both classical and AI-based control strategies. This makes the framework well suited for iterative experimentation and gradual integration of AI-driven control loops alongside proven algorithms.

\subsection{Dataset Collection and Experiment Instrumentation}

Beyond active control, the AI-NNC can be used as a unified data collection framework for testbed experiments. Since all registered nodes publish measurements through a common message broker and registration plane, the framework can be configured to collect performance metrics without issuing any control actions.

In this mode, control applications act purely as data consumers, subscribing to selected measurements and exporting them to external storage such as databases, log files, or structured formats including CSV. This enables consistent dataset generation across multi-domain testbeds, with synchronized measurements collected from all participating nodes through a uniform interface.

\subsection{Agentic AI Experimentation Interface}

The AI-NNC also provides a natural experimentation platform for agentic AI applied to network systems. Through the Model Context Protocol, large language model agents can interact with the network by invoking validated control applications, querying live measurements, and reasoning over network state.

This use case does not require additional changes to the framework. Experimenters define one or more control applications, specify safety constraints through command validators, and then connect any LLM-based agent to the MCP server. The agent can then explore agentic decision-making strategies while remaining constrained by domain-specific validation logic, making the framework suitable for controlled experimentation with autonomous AI behavior on real or emulated network infrastructure.

\subsection{Testbed Resource Coordination and Experiment Orchestration}

In a testbed environment, the AI-NNC can be used to coordinate and orchestrate concurrent experiments across shared network resources. Since the registration plane tracks which control applications are active, which nodes they control, and which control functions they invoke, it provides visibility into current resource usage.

This information can be leveraged to determine which nodes are available, which are already allocated to experiments, and how control responsibilities are distributed. In this context, control applications act as experiment orchestration scripts rather than continuous controllers. They define which nodes are involved in an experiment, configure initial conditions through validated control actions, and monitor measurements during execution.

By reusing the same registration, measurement, and validation mechanisms, the framework supports reproducible and conflict-aware experiment coordination without introducing separate testbed management infrastructure.

\subsection{Bridging Real Infrastructure with Simulators and Digital Twins}

The AI-NNC can also serve as a unified bridge between real network infrastructure and simulated environments or digital twins. Since the framework integrates any system through the same lightweight node adapter abstraction, both physical devices and simulated components can be represented uniformly as registered network nodes. In this setting, a simulator or digital twin can be integrated as a network node, or as a collection of nodes, that publishes synthetic measurements and accepts validated control commands in the same manner as real equipment.

This capability enables hybrid experimental workflows in which real network devices operate alongside their digital counterparts, allowing operators and researchers to evaluate control strategies under realistic conditions while leveraging simulation-based scalability and repeatability. Control applications and AI agents can interact seamlessly with mixed deployments, treating real and virtual nodes through the same measurement and control interfaces. As a result, the AI-NNC provides a practical foundation for digital-twin-assisted network experimentation, validation of autonomous control policies, and safe transition from simulated evaluation to deployment on physical infrastructure.

\subsection{Extensibility to Additional Research Scenarios}

The flexibility of the AI-NNC architecture enables additional research-driven use cases, including hybrid control strategies where AI agents coordinate multiple classical controllers, cross-domain optimization experiments spanning radio and transport layers, and comparative evaluation of different control policies operating on identical measurement streams.

These use cases are enabled by the separation of concerns inherent in the framework design, where node integration, control logic, AI interaction, and safety enforcement remain modular and independently extensible.
%======================================================================
\section{Conclusion}
\label{sec:conclusion}
%======================================================================

We presented the AI-Native Network Controller, a modular framework for safe agentic control of multi-domain network infrastructure. The framework's protocol-agnostic design enables integration of any network device through a single Python adapter, while its control application architecture provides validated tools for both classical and AI-based control. The command validation mechanism ensures that AI agents, operating through the Model Context Protocol, cannot bypass safety constraints, thereby providing the guarantees required for autonomous management of critical radio, optical, and core network infrastructure. The architecture supports a fundamentally safer paradigm for agentic AI in telecommunications, in which AI acts as a meta-controller that orchestrates validated domain-specific applications with full visibility into the control topology for conflict detection and resolution. The framework is open-source and has been validated using both simulated and physical optical network equipment.

Future work includes extending the conflict detection capabilities through formal verification of safety properties, integrating with O-RAN RIC interfaces for radio domain control, and evaluating the framework's performance in large-scale multi-domain deployments.

\bibliographystyle{IEEEtran}
\bibliography{references}

@article{polese2023oran,
  author  = {M. Polese and L. Bonati and S. D'Oro and S. Basagni and T. Melodia},
  title   = {Understanding {O-RAN}: Architecture, Interfaces, Algorithms, Security, and Research Challenges},
  journal = {IEEE Communications Surveys \& Tutorials},
  volume  = {25},
  number  = {2},
  pages   = {1376--1411},
  year    = {2023}
}

@article{bonati2021intelligence,
  author  = {L. Bonati and S. D'Oro and M. Polese and S. Basagni and T. Melodia},
  title   = {Intelligence and Learning in {O-RAN} for Data-Driven {NextG} Cellular Networks},
  journal = {IEEE Communications Magazine},
  volume  = {59},
  number  = {10},
  pages   = {21--27},
  year    = {2021}
}

@article{santos2025managing,
  title={Managing O-RAN networks: xApp development from zero to hero},
  author={Santos, Joao F and Huff, Alexandre and Campos, Daniel and Cardoso, Kleber V and Both, Cristiano B and DaSilva, Luiz A},
  journal={IEEE Communications Surveys \& Tutorials},
  year={2025},
  publisher={IEEE}
}

@article{foukas2017slicing,
  author  = {X. Foukas and G. Patounas and A. Elmokashfi and M. K. Marina},
  title   = {Network Slicing in {5G}: Survey and Challenges},
  journal = {IEEE Communications Magazine},
  volume  = {55},
  number  = {5},
  pages   = {94--100},
  year    = {2017}
}

@article{afolabi2018slicing,
  author  = {I. Afolabi and T. Taleb and K. Samdanis and A. Ksentini and H. Flinck},
  title   = {Network Slicing and Softwarization: A Survey on Principles, Enabling Technologies, and Solutions},
  journal = {IEEE Communications Surveys \& Tutorials},
  volume  = {20},
  number  = {3},
  pages   = {2429--2453},
  year    = {2018}
}

@article{ebrahimi2024resource,
  title={Resource management from single-domain 5g to end-to-end 6g network slicing: A survey},
  author={Ebrahimi, Sina and Bouali, Faouzi and Haas, Olivier CL},
  journal={IEEE Communications Surveys \& Tutorials},
  volume={26},
  number={4},
  pages={2836--2866},
  year={2024},
  publisher={IEEE}
}

@article{mckeown2008openflow,
  author  = {N. McKeown and T. Anderson and H. Balakrishnan and G. Parulkar and L. Peterson and J. Rexford and S. Shenker and J. Turner},
  title   = {{OpenFlow}: Enabling Innovation in Campus Networks},
  journal = {ACM SIGCOMM Computer Communication Review},
  volume  = {38},
  number  = {2},
  pages   = {69--74},
  year    = {2008}
}

@article{kreutz2015sdn,
  author  = {D. Kreutz and F. M. V. Ramos and P. Esteves Verissimo and C. Esteve Rothenberg and S. Azodolmolky and S. Uhlig},
  title   = {Software-Defined Networking: A Comprehensive Survey},
  journal = {Proceedings of the IEEE},
  volume  = {103},
  number  = {1},
  pages   = {14--76},
  year    = {2014}
}

@inproceedings{berde2014onos,
  author    = {P. Berde and M. Gerola and J. Hart and Y. Higuchi and M. Kobayashi and T. Koide and B. Lantz and B. O'Connor and P. Radoslavov and W. Snow and G. M. Parulkar},
  title     = {{ONOS}: Towards an Open, Distributed {SDN} {OS}},
  booktitle = {Proc. 3rd ACM Workshop on Hot Topics in Software Defined Networking (HotSDN)},
  year      = {2014}
}

@inproceedings{medved2014opendaylight,
  author    = {J. Medved and R. Varga and A. Tkacik and K. Gray},
  title     = {{OpenDaylight}: Towards a Model-Driven {SDN} Controller Architecture},
  booktitle = {Proc. IEEE International Symposium on a World of Wireless, Mobile and Multimedia Networks (WoWMoM)},
  year      = {2014}
}

@inproceedings{mao2016resource,
  author    = {H. Mao and M. Alizadeh and I. Menache and S. Kandula},
  title     = {Resource Management with Deep Reinforcement Learning},
  booktitle = {Proc. 15th ACM Workshop on Hot Topics in Networks (HotNets)},
  pages     = {50--56},
  year      = {2016}
}

@article{zhou2024llm,
  author  = {Z. Zhou and others},
  title   = {Large Language Model ({LLM}) for Telecommunications: A Comprehensive Survey on Principles, Key Techniques, and Opportunities},
  journal = {IEEE Communications Surveys \& Tutorials},
  year    = {2024}
}

@article{hong2025llm,
  author  = {S. Hong and others},
  title   = {A Comprehensive Survey on {LLM}-Based Network Management and Operations},
  journal = {International Journal of Network Management},
  volume  = {35},
  year    = {2025}
}

@misc{3gpp23288,
  author = {{3GPP}},
  title  = {{TS} 23.288: Architecture Enhancements for {5G} System ({5GS}) to Support Network Data Analytics Services},
  year   = {2023}
}

@misc{etsizsm002,
  author = {{ETSI}},
  title  = {{GS ZSM} 002: Zero-touch Network and Service Management ({ZSM}); Reference Architecture},
  year   = {2019}
}

@article{li2025agentic,
  author  = {X. Li and W. Shi and H. Zhang and C. Peng and S. Wu and W. Tong},
  title   = {The Agentic-{AI} Core: An {AI}-Empowered, Mission-Oriented Core Network for Next-Generation Mobile Telecommunications},
  journal = {Engineering},
  year    = {2025},
  publisher = {Elsevier}
}

@article{demirel2026intents,
  title={From Intents to Actions: Agentic AI in Autonomous Networks},
  author={Demirel, Burak and Soldati, Pablo and Wang, Yu},
  journal={arXiv preprint arXiv:2602.01271},
  year={2026}
}

@article{abou2025agentic,
  title={Agentic AI: a comprehensive survey of architectures, applications, and future directions},
  author={Abou Ali, Mohamad and Dornaika, Fadi and Charafeddine, Jinan},
  journal={Artificial Intelligence Review},
  volume={59},
  number={1},
  pages={11},
  year={2025},
  publisher={Springer}
}

@article{garcia2015safe,
  author  = {J. Garcia and F. Fernandez},
  title   = {A Comprehensive Survey on Safe Reinforcement Learning},
  journal = {Journal of Machine Learning Research},
  volume  = {16},
  pages   = {1437--1480},
  year    = {2015}
}

@inproceedings{alshiekh2018shielding,
  author    = {A. Alshiekh and R. Bloem and R. Ehlers and B. K{\"o}nighofer and S. Niekum and U. Topcu},
  title     = {Safe Reinforcement Learning via Shielding},
  booktitle = {Proc. AAAI Conference on Artificial Intelligence},
  volume    = {32},
  year      = {2018}
}

@misc{nist2023ai,
  author = {{NIST}},
  title  = {Artificial Intelligence Risk Management Framework ({AI RMF} 1.0)},
  year   = {2023},
  note   = {NIST AI 100-1}
}

@misc{dhs2024ai,
  author = {{U.S. Department of Homeland Security}},
  title  = {Safety and Security Guidelines for Critical Infrastructure Owners and Operators: {AI} in Critical Infrastructure},
  year   = {2024}
}

@inproceedings{dzaferagic2025icc,
  author    = {M. Dzaferagic and M. Ruffini and N. Slamnik-Krijestorac and J. F. Santos and J. Marquez-Barja and C. Tranoris and S. Denazis and G. C. Tziavas and T. Kyriakakis and P. Karafotis and L. A. {DaSilva} and S. R. Pandey and J. Shiraishi and P. Popovski and S. K. Jensen and C. Thomsen and T. B. Pedersen and H. Claussen and J. Du and G. Zussman and T. Chen and Y. Chen and S. Tirupathi and I. Seskar and D. Kilper},
  title     = {Decentralized {AI}-Control Framework for Multi-Party Multi-Network {6G} Deployments},
  booktitle = {2025 IEEE International Conference on Communications Workshops (ICC Workshops)},
  pages     = {1227--1232},
  year      = {2025},
  address   = {Montreal, Canada}
}

@article{dzaferagic2025modular,
  author  = {M. Dzaferagic and M. Ruffini and D. Kilper},
  title   = {Modular and Integrated {AI} Control Framework across Fiber and Wireless Networks for {6G}},
  journal = {arXiv preprint arXiv:2502.15731},
  year    = {2025}
}

@misc{zmq,
  author = {{iMatix Corporation}},
  title  = {{ZeroMQ}: Distributed Messaging},
  howpublished = {\url{https://zeromq.org}},
  year   = {2024}
}

\end{document}